\documentclass[preprint2]{emulateapj}

\shorttitle{{\sl Suzaku} observations of V2491 Cyg}
\shortauthors{Zemko et al.}

\begin{document}

\title{{\sl Suzaku} observation of the classical nova V2491 Cyg in quiescence.}

\author{P. Zemko} 
\affil{Department of Physics and Astronomy, Universit\`a di Padova, vicolo dell' 
Osservatorio 3, I-35122 Padova, Italy}
\email{polina.zemko@studenti.unipd.it}

\author{K. Mukai\altaffilmark{1}}
\affil{CRESST and X-ray Astrophysics Laboratory, NASA Goddard Space Flight Center, 
Greenbelt, MD 20771, USA }
\email{koji.mukai@umbc.edu}
\and
\author{ M. Orio\altaffilmark{2}}
\affil{INAF - Osservatorio di Padova, vicolo dell' Osservatorio 5, I-35122 Padova, Italy}
\email{marina.orio@oapd.inaf.it}

\altaffiltext{1}{Department of Physics, University of Maryland, Baltimore County, 1000 
Hilltop Circle, Baltimore, MD 21250, USA}
\altaffiltext{2}{Department of Astronomy, University of Wisconsin, 475 N. Charter Str., 
Madison, WI 53704, USA}

\begin{abstract}
We present {\sl Suzaku} XIS observation of V2491 Cyg (Nova Cyg 2008 No. 2)
 obtained in quiescence, more than two years after the outburst. The nova was detected as
 a very luminous source in a wide spectral range from soft to hard X-rays. A very soft 
 blackbody-like component peaking at 0.5 keV indicates that either we observe remaining,
  localized hydrogen burning on the surface of the white dwarf, or accretion onto a
   magnetized polar cap. In the second case, V2491 Cyg is a candidate ``soft 
   intermediate polar''. We obtained the best fit for the X-ray spectra with several 
   components: two of thermal plasma, a blackbody and a complex absorber. The later 
   is typical of intermediate polars.
   The X-ray light-curve shows a modulation with a $\sim$38 min period. The 
 amplitude of this modulation is strongly energy dependent and reaches maximum in the 
 0.8--2.0 keV range. 
We discuss the origin of the X-ray emission and pulsations, and the likelihood of the 
intermediate polar scenario.
\end{abstract}


\section{Introduction}

Cataclysmic variables are close binary systems consisting of a white dwarf (WD)
primary and a low-mass late type main sequence star transferring material via Roche-lobe 
overflow. If enough material is accumulated, the temperature and the pressure at the bottom 
of the accreted envelope can be high enough for a termonuclear runaway and
a classical nova (CN) explosion \citep[see e.g.][]{sta12,wol13Hburn}. 
When the ejected envelope becomes transparent to soft X-rays the system is observed as 
a super-soft X-ray source (SSS) revealing the hot WD atmosphere and nuclear 
burning \citep[e.g.][]{ori01rosat, kra08sss, ori12gratings}.

V2491 Cyg was detected by Hiroshi Kaneda on 2008 April 10.728 at $V$=7.7 
\citep{nak08V2491cyg}. The nova was very fast: $t_2$ (time needed to decrease in 
luminosity by 2 mag) in the $V$ band was 
4.6 days and the ejecta velocity reached 4860 km s$^{-1}$ \citep{tom08V2491Cyg2, tom08V2491Cyg1}. 
The nova had a short unusual ``re-brightening'' at the end of April, then 
reached the minimum brightness -- $V$ $\sim $16 after about 150 days. The interstellar 
reddening was E(B-V) = 0.43 \citep{rud08V2491cyg} corresponding to a hydrogen column 
 N(H)=2.5$\times10^{21}$ cm$^{-2}$ \citep{boh78abs}.

\citet{jur08V2491Cyg} discovered a persistent optical counterpart at V$\simeq$17.1, which had a 
short dimming before the nova outburst.
No previous outburst was known, but several authors suggested that an early transition of 
the optical spectrum to the He/N type \citep{tom08V2491Cyg2}, the velocity of the ejecta, and 
the rapid visual decay were all typical of the recurrent novae (RNe). However, this hypothesis
 has not been proven.

V2491 Cyg is the second out of three exampless of CNe detected in X-rays before the outbursts
(see \citealt{her02V2487Ohp} for V2487 Oph and \citealt{sch08N2008Car} for Nova 2008 Car).
Pre-nova X-ray observations of V2491 Cyg obtained with {\sl ROSAT}, {\sl XMM-Newton}, and 
{\sl Swift} were discussed in \citet{iba09V2491cyg}. These authors found that the quiescence 
X-ray spectrum was variable on a time scale as short as 4 days and that one of the analysed
 {\sl Swift} spectra was noticeably softer than at other epochs. The unabsorbed X-ray
 flux varied in the range from 1 to 30$\times 10^{-12}$ ergs cm$^{-2}$ s$^{-1}$ \citep{iba09V2491cyg},
  corresponding to $L_{\rm{X}}=1.2-4\times 10^{35}$ erg s$^{-1}$, assuming a distance of
10.5 kpc \citep{hel08V2491cyg}. These values of the X-ray luminosity before the outbursts imply
that the mass accretion rate can be as high as $\sim $10$^{-8}$ M$_{\sun}$ yr$^{-1}$ for a
 1.3 M$_{\sun}$ WD \citep{hac09V2491Cyg}, close indeed to the expected RN range, with 
 recurrence times $\sim$100 years \citep{wol13Hburn}.

During the outburst V2491 Cyg was observed with {\sl XMM-Newton} 
\citep{nes11V2491Cyg, tak11V2491Cyg}, {\sl Suzaku} \citep{tak11V2491Cyg} and extensively 
monitored with {\sl Swift} \citep{kuu08V2491Cyg, osb08V2491Cyg, pag10V2491cyg}.
\citet{pag10V2491cyg} followed V2491 Cyg with {\sl Swift}, from the day 
after the nova discovery until the pre-outburst flux level. The spectrum was quite hard 
before day 25 after the explosion and shortly thereafter the object evolved into
 a SSS. The peak soft X-ray luminosity remained constant for only two days and than  
faded slowly for 18 days, an unsual trend for post-nova SSS, most of which show an 
almost flat light curve after the peak and later fade rapidly.
The WD in V2491 Cyg was among the hottest ever observed \citep{nes11V2491Cyg}, therefore it must be 
very massive \citep{sta12,wol13Hburn}. Quite surprisingly, the WD did not seem to cool 
before the SSS final decay, while the luminosity significantly decreased, as if the nuclear
 burning region was shrinking on the surface of the WD itself, at constant temperature 
 (most nova SSS are observed to cool, see e.g. the review by \citealt{ori12gratings}). 
 
\citet{bak08V2491Cyg} reported a variation with a period of 0.09580(5) days 
in the $B$ and $V$ bands between 10 and 20 days after the outburst.
This variation may have been the orbital modulation of the binary system. 
However, \citet{shu10v2491cyg} did not detect the above period, although they monitored V2491 Cyg
 in optical for more than one year. Also \citet{dar11v2491cyg}
ruled out eclipsing orbital periods shorter than 0.15 days.
\citet{pag10V2491cyg} observed V2491 Cyg during the decline to quiescence, but
did not detect any modulation with a period $\sim$ 0.1 days in X-rays and UV.
While the 0.09580(5) days period has not been confirmed, another variability on a shorter
timescale has been reported by several authors. \citet{shu10v2491cyg}
detected a possible 0.02885 days (41 min) period. \citet{dar11v2491cyg} found evidence 
of a $\sim$0.025 days (36 min) modulation in the $B$ band, but, unfortunately, their data were too 
sparse for period analysis. \citet{nes11V2491Cyg} reported oscillations of the X-ray flux 
with a period of 37.2 min on day 39 after the outburst, however this variability was 
not observed in simultaneous UV observations and in the X-ray light curve obtained
later, on day 49 after the nova explosion. 
 
Several authors discussed the possibility of a magnetic scenario for V2491 Cyg
\citep[see][]{hac09V2491Cyg, iba09V2491cyg, tak11V2491Cyg}. \citet{hac09V2491Cyg} proposed 
that magnetic activity explains the re-brightening seen in the optical light curve, and
that V2491 Cyg is a polar. However, \citet{pag10V2491cyg} argued against this 
possibility. Using the synchronization condition they showed that if V2491 Cyg is a polar
 it should host one of most magnetic WDs known in binaries, assuming the WD mass about 
 1.3$M_{\sun}$ and 0.0958 days 
orbital period. Although V2491 Cyg has a number of properties that are quite typical of 
magnetic WDs, such as strong Fe emission features at 6.4, 6.7, and 7.0 keV, high X-ray 
luminosity in the range 2.0–-10 keV \citep{tak11V2491Cyg}, 
the existing data do not allow to finally prove or disprove the magnetic scenario.
 
 The indications of large WD mass and high $\dot{m}$ suggest that this nova is a rather 
``extreme'' accreting WD, with characteristics we expect in a type Ia supernova (SN) progenitor 
and possibly with a strong magnetic field. We observed this intriguing CN with the
{\sl Suzaku} X-ray Imaging Spectrometer (XIS) in quiescence in order to monitor wether if the WD is 
still hot and to shed some light on its possible magnetic nature.

\section{Observations and data analysis}

V2491 Cyg was observed with the {\sl Suzaku} XIS on 2010 November 3 with an exposure time 
of 74.4 ks (for details see Table 1). The data were processed and analysed using HEASOFT v.6-13.
We read the event files with XSELECT v2.4b and generated the detector and
mirror responses using {\it xisarfgen} and {\it xisrmfgen} tools. We 
combined the XIS 0 and XIS 3 data 
from the front-illuminated (FI) CCD chips using 
{\it addascaspec} and the XIS 1 data from the back-illuminated (BI) CCD were analysed separately.
The spectral analysis was performed with XSPEC v.12.8.0. The X-ray light curves
 were extracted after the barycentric correction and processed with XRONOS package.

\begin{table*}
\clearpage
 \centering
 \begin{minipage}{120mm}
  \caption{Observational log of the {\sl Suzaku} observations of V2491 Cyg.}
  \begin{tabular}{lccc}
\tableline
\tableline
Date and time & Instrument & Exposure (s) & Count rate (cts s$^{-1}$) \\
\tableline
2010-11-03 10:32:11  & {\sl Suzaku} XIS 0 & 74400  & $0.0494\pm0.0010$\\  
2010-11-03 10:32:11  & {\sl Suzaku} XIS 1 & 74400  & $0.0681\pm0.0012$\\  
2010-11-03 10:32:11  & {\sl Suzaku} XIS 3 & 74400  & $0.0550\pm0.0010$\\ 
\tableline
\multicolumn{4}{p{.9 \textwidth}}{{\textbf Notes.} {\it Suzaku} XIS FI are XIS 0 and XIS 3
detectors with front-illuminated (FI) CCDs, while {\it Suzaku} XIS BI is the XIS 1 that 
utilizes a back-illuminated (BI) CCD}\\
\end{tabular}
\end{minipage}
\label{tab:obs}
\end{table*}

\section{Results}\label{res}

\subsection{Spectral analysis}\label{spec}
\begin{figure*}
\centering
\includegraphics[angle=270, width=320pt]{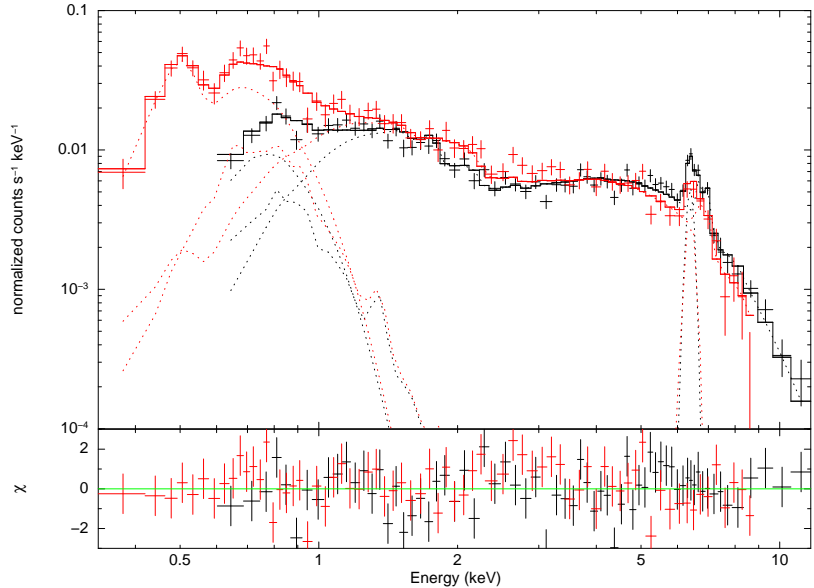}
\caption{{\sl Suzaku} XIS spectra of V2491 Cyg. The BI CCD data have been plotted in red and 
the FI data have been
plotted in black. The solid lines represent the fit with wabs$\times$(bb+pcfabs$\times$(apec+apec+gauss))
model. The components of the model have been plotted with the dashed lines.}
\label{fig:spectra}
\end{figure*}

The background subtracted 0.3--10.0 keV spectra of V2491 Cyg are presented in Figure~
\ref{fig:spectra}. The combined XIS 0 and the XIS 3 data are plotted 
in black, while the XIS 1 data are plotted in red. Solid lines show the best fit. The 
dashed lines represent the components of this model. The spectra seem
 to have a very soft component and a harder one with emission lines of highly ionized Fe, 
 in particular the \ion{Fe}{25} line at 6.7 keV that indicates a thermal plasma. In order to fit 
the harder portion of the spectra we started with one thermal plasma component. However neither one,
nor two components of collisionally-ionized diffuse gas ({\sl APEC} model in 
XSPEC, calculated using the ATOMDB code v2.0.2) did not provide a statistically 
significant fit of the spectra with any temperature. The multi-temperature plasma emission model based on the 
{\sl MEKAL} code ({\it CEMEKL}) also does not provide a good fit. We tried to add 
a reflection component, taking into account the presence of the very strong Fe K$\alpha$ 
reflection line, but it resulted in unphysical values of the 
 fitting parameters. The reflection scaling factor (R$=\Omega/2\pi$), which is the 
 covering fraction of the reflector viewed from the plasma, was close to 10. Such a value of
 R would imply that the source of the X-ray emission is hidden by a Compton-thick material
  and that we only see the reflected component, which 
is a rather unlikely possibility. Moreover, R$=10$ implies that the EW of the Fe K$\alpha$
 reflection line should be $\approx$1.5 keV \citep{geo91FeK}, which is much larger than the 
EW inferred from the Gaussian fit of the line that we will show below.

The very flat slope in the 3.0--5.0 keV range points toward a complex absorption. 
The fit with the two-component thermal plasma model was much 
 improved after multiplication with a partially covering absorber ({\it PCFABS} in 
 XSPEC). We derived the following values of the column density
 and covering fraction of the partially covering absorber: N(H) = 10--17 $\times$ 10$^{22}$ cm$^{-2}$ and
CvrFract = 0.66. \citet{pag10V2491cyg} also used a partially covering absorber with 
 almost the same value of N$_H$ in their model to fit the spectra of V2491 Cyg obtained 
 after day 80. 

 In order to fit the soft excess, seen in the XIS 1 data at 0.5 keV, we 
 added a blackbody component that was absorbed only by the simple absorption. We will 
 further discuss why the blackbody component is not effected by the 
 partially covering absorber. By adding to this four components a Gaussian to
 represent the Fe K$\alpha$ line at 6.4 keV, we finally obtained a statistically acceptable result
  with a value of $\chi^2$ equal to 1.1. The abundance of the thermal plasma components
  is 0.63$_{-0.20}^{+0.27}$ with respect to solar. 
The value of N(H) derived from our fit is in agreement 
 with the pre-outburst values reported by \citet{iba09V2491cyg} and with the estimates 
 based on the interstellar reddening. The components and parameters of 
 the fit are presented in Table 2.

\begin{table}
\clearpage
\centering
\begin{minipage}{80mm}
  \caption{The parameters of the best fitting models -- 
  wabs$\times$(bb/``{\it atm}''+pcfabs$\times$(apec+apec+gauss)) for V2491 Cyg X-ray spectra, where 
  ``{\it atm}'' is the WD atmosphere model.}
   \begin{tabular}{l cc}
\tableline
\tableline
Parameter								& +Blackbody				& +Atmosphere  \\
\tableline
N(H)$^{[1]}$ 							& $0.25_{-0.06}^{+0.08}$ 	&$0.19_{-0.07}^{+0.09}$\\
N(H)$_{\rm{pc}}$$^{[2]}$ 				& $13.3_{-2.4}^{+3}$  		&$10.8_{-2.0}^{+2.5}$\\
CvrFract$^{[3]}$ 						& $0.66_{-0.03}^{+0.03}$	&$0.63_{-0.03}^{+0.03}$\\
T$_{\rm bb/atm}$ (eV)					& $77_{-9}^{+7}$			&$85.9_{-1.5}^{+2.2}$\\
T$_1$ (keV)								& $0.24_{-0.24}^{+0.24}$	&$0.29_{-0.03}^{+0.05}$\\
T$_2$ (keV)								& $11.3_{-1.5}^{+1.8}$		&$15.0_{-2.4}^{+4.3}$\\
Norm$_1$$^{[4]}$						& $5_{-5}^{+8}$				&$6.9_{-2.8}^{+4.6}$\\
Norm$_2$$^{[4]}$						& $14.6_{-1.0}^{+1.1}$		&$14.1_{-1.0}^{+1.1}$\\
Flux$_{\rm{abs}}$$^{[5]}$ 				& $1.92_{-0.23}^{+0.03}$  	&$1.93_{-0.6}^{+0.11}$ \\
Flux$_{\rm{unabs}}$$^{[5]}$ 			& 6.66 						&3.95 \\
Flux$_{\rm{abs}}^{\rm{bb}/atm}$$^{[6]}$ & 0.346  					&0.284\\
Flux$_{\rm{unabs}}^{\rm{bb}/atm}$$^{[6]}$& 7.80	  					&2.66\\
$\chi^2$								& 1.1	 	  				&1.1\\ 
L$_{\rm{2.0-10.0 keV}}$$^{[7]}$ 		& 1.82						&1.82\\
L$_{\rm bb/atm}$$^{[8]}$ 				& $1.4_{-0.7}^{+2.4}$		&$1.72_{-0.9}^{+2.9}$\\
R$_{\rm bb/atm}$ cm$^{[9]}$				& $1.2_{-0.4}^{+1.0}$		&$1.4_{-0.5}^{+0.9}$ \\

\tableline 
\multicolumn{3}{p{.9\textwidth}}{{\textbf Notes.} The errors represent the 90\% confidence region for a single parameter.} \\
\multicolumn{3}{p{.9\textwidth}}{$^{[1]}$ $\times10^{22}$ cm$^{-2}$ }\\
\multicolumn{3}{p{.9\textwidth}}{$^{[2]}$ $\times10^{22}$ cm$^{-2}$, N(H) of the partial covering absorber.}\\
\multicolumn{3}{p{.9\textwidth}}{$^{[3]}$ Covering fraction of the partial covering absorber.}\\
\multicolumn{3}{p{.9\textwidth}}{$^{[4]}$ Normalization constants of the apec model ($\times10^{-4}$ cm$^{-5}$).}\\
\multicolumn{3}{p{.9\textwidth}}{$^{[5]}$ The X-ray flux ($\times10^{-12}$ erg cm$^{-2}$ s$^{-1}$) 
measured in the range 0.3--10.0 keV. Flux$_{\rm{unabs}}$ represents the value of the X-ray flux, 
corrected for the interstellar absorption only.}\\
\multicolumn{3}{p{.9\textwidth}}{$^{[6]}$ The X-ray flux ($\times10^{-12}$erg cm$^{-2}$ s$^{-1}$) of the 
blackbody and WD atmosphere components measured in the range 0.2--10.0 keV.}\\
\multicolumn{3}{p{.9\textwidth}}{$^{[7]}$ The X-ray luminosity ($\times10^{34}$ erg s$^{-1}$ D$_{10.5 kpc}^2$). 
We assumed D=10.5 kpc \citep{hel08V2491cyg}.}\\
\multicolumn{3}{p{.9\textwidth}}{$^{[8]}$ The bolometric X-ray luminosity ($\times10^{35}$ 
erg s$^{-1}$ D$_{10.5 kpc}^2$) of the blackbody and atmospheric components. L$_{{\rm bb}/atm}$ were
 calculated based on the normalization constants of the models. 
The atmospheric model gives the value of the WD radius R$_{\rm WD}=10^{-11}\times\sqrt{{\rm norm}}\times$D, 
which can be translated to the L$_{atm}$ using the Stefan-Boltzmann law.}\\
\multicolumn{3}{p{.9\textwidth}}{$^{[9]}$ The radius of the emitting region ($\times10^7$ cm D$_{10.5 kpc}$)}\\

\end{tabular}
\end{minipage}
\label{tab:fit}
\end{table}
 
The origin of the blackbody-like component in the spectra of V2491 Cyg is not quite clear. 
\citet{pag10V2491cyg} speculated about the possibility of long-lasting hydrogen burning on
 the surface of the WD in this system. Aiming at distinguishing the possible sources of the 
 soft X-ray emission in V2491 Cyg we compared the blackbody model with a WD 
 atmosphere model, since the latter better describes the hydrogen burning on a WD.
 We used in XSPEC the publicly available tabulated T\"ubingen Non Local Thermal Equilibrium 
 Atmosphere Model (NLTE TMAP) for a WD atmosphere in spherical or plane-parallel 
 geometry in hydrostatic and radiative equilibrium, described by \citet{rau03NLTE}, with
chemical composition of elements from H to Ni $\#$007 and $\log{g}=9$. The blackbody and the NLTE 
TMAP are statistically equal, -- both give the value of $\chi^2=1.1$. The values
 of the temperature and luminosity, and hence the emitting area, derived from the blackbody 
 and WD atmosphere model are comparable (see Tab. 2).

\begin{figure}
\includegraphics[angle=270, width=210pt]{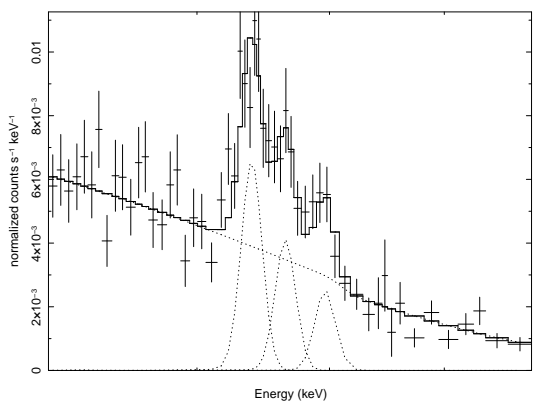}
\caption{The {\sl Suzaku} XIS spectra of V2491 Cyg in the 5.0--9.0 keV range showing the Fe K complex.
 The three Gaussians represent the \ion{Fe}{1}, \ion{Fe}{25} and \ion{Fe}{26} lines.}
\label{fig:iron}
\end{figure}

Finally we isolated the 5.0--9.0 keV region and for simplicity fitted it  with a 
power-law continuum with $\Gamma=2.0$ and three
 Gaussians representing the K$\alpha$ fluorescent line of \ion{Fe}{1} (corresponding to
 2{\rm p} $\rightarrow$ 1{\rm s} electron transition), a \ion{Fe}{25} 
 resonance line and a Ly$\alpha$ line of \ion{Fe}{26} (see Figure~\ref{fig:iron}). 
 The N(H) and parameters of the partially covering absorber were fixed to the 
 values of the best-fitting model in Table 2.  
 First, we fixed the centroids of the lines to the rest values and the Gaussian widths to zero, 
 since the natural widths of the lines 
 and the broadening due to thermal motions of the emitting atoms are negligible compared 
 with the instrumental resolution of the {\sl Suzaku} XIS detectors, which is $\sim$130 eV at 6 keV. 
 This fit does not provide a good result, leaving a residual excess between the 6.4 
and 6.7 keV lines. This may be due to the complex structure of the \ion{Fe}{25} feature: 
it consists of the resonance line, the forbidden line, and two intercombination lines. 
Moreover, there are dielectric satellite lines in the range of 6.61--6.68 keV 
\citep{hel04FeK}. Therefore, we varied the centroid position of \ion{Fe}{25} feature and 
significantly improved the fit. The best fitting parameters for the Fe K complex and the 
EWs of the lines are presented in Table 3. The EW of the Fe K$\alpha$ line in the {\sl Suzaku}
spectra is $\sim$246 eV, which is comparable with the value measured on day 60 and 150 after
 the outburst \citep{tak11V2491Cyg}.
 
  The flux ratio of the \ion{Fe}{26} and \ion{Fe}{25} lines gives an estimate 
  of the highest temperature in the post-shock region \citep{sch14lineratios}. From the fit
   of the Fe K$\alpha$ complex of V2491 Cyg we 
 found that this ratio is $\sim$0.7 (see Table 3), which corresponds to a 
 ionization temperature of about 10 keV \citep{sch14lineratios}. This value is close to 
 that derived from the global fit (see Table 2).

\begin{table*}
\clearpage 
\centering
\begin{minipage}{90mm}
  \caption{The best-fit parameter values for the Fe K lines fitting.}
  \begin{tabular}{l|ccc}
\tableline
\tableline
Parameter					&	\ion{Fe}{1} 			& \ion{Fe}{25}			 & \ion{Fe}{26}  \\
  \tableline
Energy center (keV) 		& $6.40$	(frozen)		& $6.65_{-0.03}^{+0.03}$ & $6.97$ (frozen)\\
EW	(eV)$^*$				& $246_{-50}^{+53}$ 		& $133_{-34}^{+48}$		 & $132_{-45}^{+56}$ \\
Flux$_{\rm{unabs}}$$^{**}$	& 7.20						& 4.95					 & 3.49\\
\tableline 
\multicolumn{4}{p{.9\textwidth}}{{\textbf Notes.} $\chi^2$ is 0.98. All the parameters were derived at 90\% confidence level.}\\
\multicolumn{4}{p{.9\textwidth}}{$^*$ Equivalents width.}\\
\multicolumn{4}{p{.9\textwidth}}{$^{**}$ The X-ray flux ($\times10^{-14}$ erg cm$^{-2}$ s$^{-1}$).}\\
\end{tabular}
\end{minipage}
\label{tab:iron}
\end{table*} 

\subsection{Timing analysis}\label{timing}

\begin{figure}
\includegraphics[width=220pt]{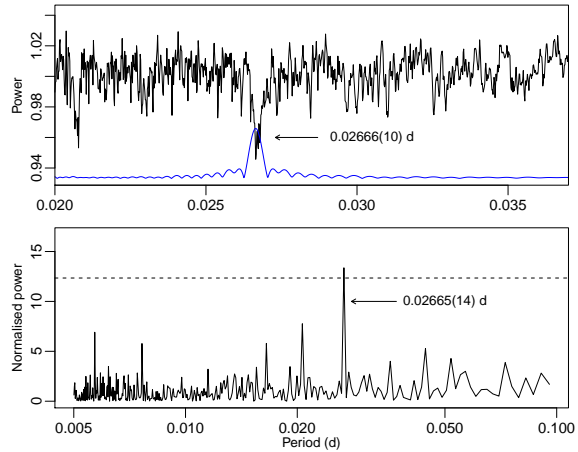}
\caption{Periodogram of the {\sl Suzaku} light curve of V2491 Cyg. We analysed the BI data
in the 0.3--10.0 keV energy range binned every 80 seconds. The upper power spectrum was obtained with the PDM method
and the bottom plot with the LS method. The highest peaks are marked with the arrows together with
the corresponding values of the periods. The false alarm probability of 0.3$\%$ level is marked
with the dashed line in the bottom plot. }
\label{fig:pdm}
\end{figure}

For the timing analysis of V2491 Cyg we initially combined the FI and BI data in the 
0.3--10 keV range and detrended the light curves with an three-order polynomial. We searched 
periodic variations using both the phase dispersion minimisation 
(PDM) method, introduced by \citet{PDM} and the Lomb-Scargle (LS) method \citep{sca82LS}. 
We found a peak at 0.02666(10) days with the PDM and at 0.02665(14) days with the LS method 
(see Figure~\ref{fig:pdm}). We did not detect any 
reliable periodic signal close to the proposed orbital period of 0.09580(5) days \citep{bak08V2491Cyg}. 
A false alarm probability level of 0.3$\%$ is marked with the horizontal line in the LS periodogram.
The peak that corresponds to the 0.0266 days period lies above this line and can be considered
 statistically significant at the 3$\sigma$ ( 99.7$\%$) level. 
 In order to further investigate the significance of the highest peak,
 we applied the bootstrap method. We repeatedly scrambled the data sequence and calculated
the probability that random peaks in the 0.005--0.1 days range reach or exceed the peak
of the unscrambled periodogram. We performed 10000 simulations and found that the 
probability that the peak is real is 99.7\%.
 
\begin{figure}
\includegraphics[width=210pt, height=260pt]{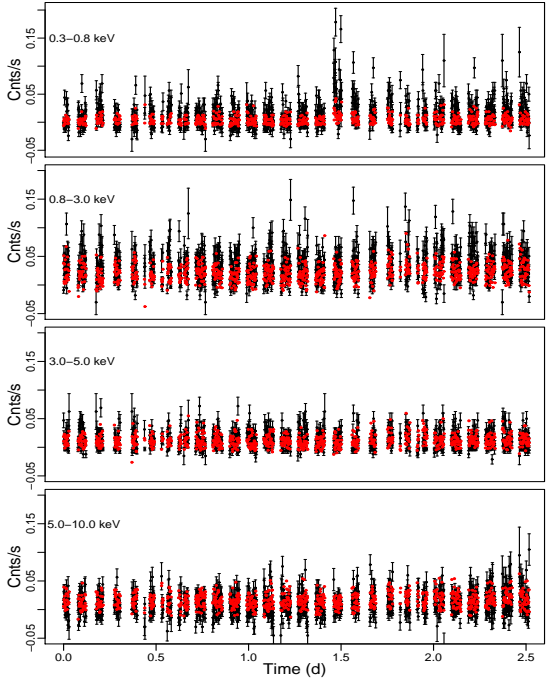}
\caption{The X-ray light curve of V2491 Cyg binned every 
80 seconds. The BI data are plotted in black and FI data - in red.}
\label{fig:xrlc}
\end{figure}

The next step was to study the energy dependence of the pulses. We extracted the light curves 
from the FI and BI data independently in the following energy ranges: 0.3--0.8 keV, 
 0.8--3.0 keV, 3.0--5.0 keV and 5.0--10.0 keV. 
The comparison of the light-curves in different ranges is presented in Figure \ref{fig:xrlc}. 
We noticed that the amplitude of 
the variations in the BI light curve is larger than in the FI data. In the 0.3--0.8 keV BI
 light curve a flare-like event can be seen, that is almost absent in the FI data in the 
 same energy range because of the low sensitivity of the FI CCD in the soft X-rays. We verified
  that this flare was not a background event. We convolved the BI light curves in 
  different energy ranges with the same period 0.0266 days
(38.3 min) and found that the amplitude of the modulation decreases with energy. The modulation is 
present only in the range below 3 keV.

\begin{figure}
\includegraphics[width=210pt, height=260pt]{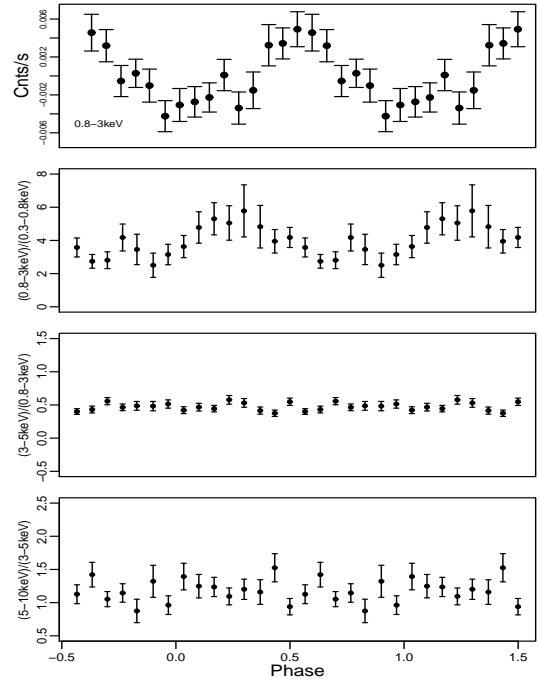}
\caption{Top panel: the {\sl Suzaku} FI+BI light curve in the range 0.8--3 keV binned every 80 s
and folded with the period of 0.0266 days. Lower panels: variations of the hardness ratios 
(HR$^i$ = N$^i$$_m$/N$^i$$_n$) in different energy ranges ($m$ and $n$) during the 0.0266 days period.}
\label{fig:hr}
\end{figure}

 We also calculated the hardness ratios HR$^i$ = N$^i$$_m$/N$^i$$_n$, where $i$ is a phase 
interval and N$^i$$_m$, N$^i$$_n$ are the number of photons in the energy ranges $m$ and $n$, 
 respectively. It can be seen from Figure \ref{fig:hr} that
 HR$^i$ = N$^i$$_{0.8-3}$/N$^i$$_{0.3-0.8}$ showed hardening at pulse minimum while 
 HR$^i$ = N$^i$$_{3-5}$/N$^i$$_{0.8-3}$ and 
 HR$^i$ = N$^i$$_{5-10}$/N$^i$$_{3-5}$ were constant within the errors.

\begin{figure}
\includegraphics[width=210pt]{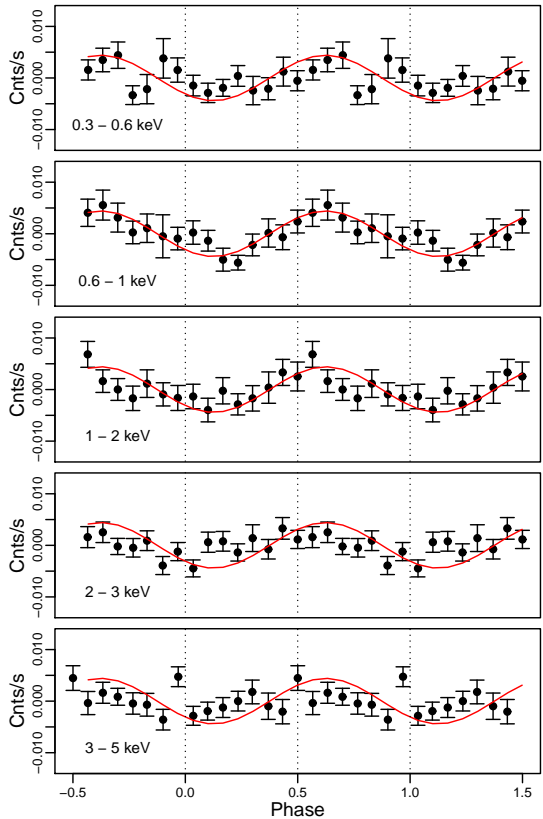}
\caption{The {\sl Suzaku} BI light curves of V2491 Cyg in different energy ranges 
folded with the period of 0.0266 days. The corresponding energy range is indicated in the 
bottom-left corner of each plot.}
\label{fig:bi30min}
\end{figure}

In order to localize the pulsating component 
we subdivided the 0.3--3.0 keV energy range in three spectral intervals: 0.3--0.6 keV, 
0.6--1.0 keV, 1.0--2.0 keV and 2.0--3.0 keV. The energy intervals were 
chosen based on the spectral fit. We expected the 0.3--0.6 keV range to be dominated 
by the blackbody, the 1.0--2.0 and 2.0--3.0 keV intervals to represent mostly the 
high temperature thermal plasma emission, and the 0.6--1.0 keV range to include emission from the 
blackbody and two thermal plasma components at the same time. 
The result can be seen in Figure ~\ref{fig:bi30min}. The periodic variation 
is observed between the  0.6--2.0 keV , but it is almost negligible in the harder ranges. 
The pulse profiles are roughly phase aligned. There is a slight shift between the maxima 
of the profiles that is within the errors. In the 0.3--0.6 keV energy range 
the variations have the same amplitude, but are irregular.

\begin{figure}
\includegraphics[width=230pt]{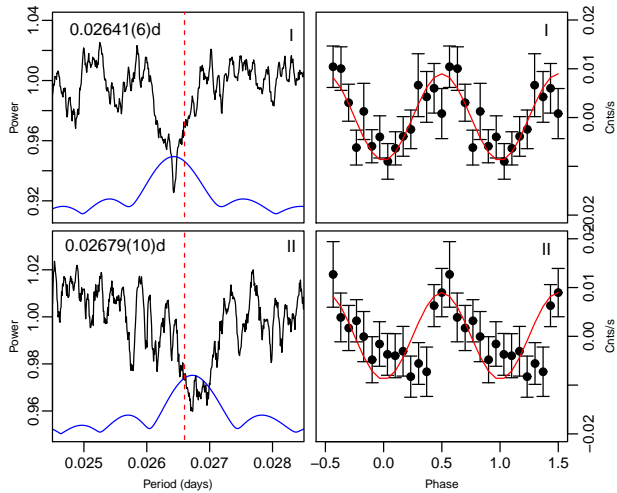}
\caption{Left panels: power spectra of the first (top) and the second (bottom) halves of 
the {\sl Suzaku} XIS BI observation of V2491 Cyg obtained with the PDM method. The red dashed line 
represent the mean period -- 0.0266 days (38.3 min). Periods measured in each half are marked. 
Right panels: the first (top) and the second (bottom) halves of the {\sl Suzaku} XIS BI light curve
folded with the 0.0266 days period (38.3 min). We analysed the light curve extracted in the 0.8--3.0 keV
energy range and binned every 80 s.}
\label{fig:halfs}
\end{figure}

In order to asses the stability of this period we split the {\sl Suzaku}
 light curve in two intervals and searched the period in each of them. In Fig. \ref{fig:halfs} 
 the periodogram (left panels) and the phase folded light curves (right panels) of the 
 first and the second parts of the dataset are plotted together. The light curves were folded 
 with the same period of 0.0266 days in order to compare the pulse profiles. We found that
  the $\sim$38 min period does not seem to be stable: 
  the best-fit periods measured in the first and the second halves of the observation are 
  different by 1.4\%.

 The energy dependence of the amplitude and the hardening at minimum suggest 
that absorption causes the observed 38.3 min modulation in the X-ray flux. The absence of 
clear modulation in the 0.3--0.6 keV range (where the blackbody emission dominates) 
indicates that the blackbody-like component is not affected by the partially covering absorber.
This is the reason why we only applied the interstellar absorption to the blackbody
component. \citet{eva07softIP} discussed the geometry which allows to observe a 
blackbody-like component in a soft IP even when the accretion curtain crosses the line of sight.

\section{Discussion}{\label{seq:disc}}

\subsection{The blackbody-like component}
Using the {\it fakeit} tool in XSPEC, we found that the count rate obtained with 
{\sl Suzaku} corresponds to a 0.039 cnts s$^{-1}$ {\sl Swift} count rate, which is 
comparable with the last {\sl Swift} observations presented in \citet{pag10V2491cyg}, 
almost 250 days after the outbursts. 
\citet{pag10V2491cyg} obtained the best fit of the X-ray spectra of
V2491 Cyg with an optically thin hard X-ray plasma and added a blackbody for the SSS phase. 
These authors also showed that the temperature of the blackbody initially increased and then 
stabilised around day 57 at the value $\sim 70$ eV. Further decrease of the X-ray 
flux in the soft X-ray range seemed to result of a shrinking of the emitting region. 
The emitting region continued to shrink even after it reached the size of the WD, instead 
of becoming cooler at a constant radius like in many other novae \citep[see][for V4743 Sgr]{rau10V4743Sgr}.

In our data we found that the blackbody-like component is still present and has a 
temperature $77_{-9}^{+7}$ eV. The normalization constant of the blackbody model gives 
a luminosity $1.4\times10^{35}$ D$_{10.5 kpc}^2$ erg s$^{-1}$ and an emitting radius 
$1.2\times10^7$ D$_{10.5 kpc}$ cm. A close emitting radius value was found in 
the fit with the stellar atmosphere model with a slightly higher temperature. Such a 
radius of the emitting region is too large for a polar cap on a magnetic WD, but is still 
more than an order of magnitude smaller than a WD radius. The fit with the WD atmosphere 
model is statistically indistinguishable
 from the one with a blackbody and does not allow to choose 
between the localized hydrogen burning and the polar cap as a possible source of the soft X-ray 
radiation. 

The constant-temperature at decreasing radius, well below the WD radius 
 dimensions, derived by \citet{pag10V2491cyg} was not observed in other novae and 
 contradicts the models. We are intrigued by the possibility that we still observe the same
  blackbody component after more than two years. Could it be due to residual nuclear burning, 
 possibly continuing in this post-outburst quiescent phase in a smaller region than the whole WD? 
 \citet{ori93local} predicted the possibility of localized thermonuclear burning on a WD. 
 Other authors have argued that the burning would not remain localized  
 and eventually the thermonuclear flash would propagate over the whole WD surface, 
 \citep[e.g.][]{gla12conv}, very differently
  for instance from helium burning on a neutron star.

On the other hand, a distinct blackbody component with temperature in the range 20--100 eV, 
heavily absorbed by dense material partially covering the X-ray source, is observed in many
 IPs \citep[and references therein]{eva07softIP, anz08softIP, ber12newIP}. These objects 
 form a growing group of so-called ``soft IPs'', including fifteen confirmed IPs
 \citep{ber12newIP}. \citet{ber12newIP} in their nine objects 
 sample found two new IPs with a blackbody component with T$_{\rm{bb}}$=70--80 eV. These 
 authors showed that the WD spot area (the source of a blackbody radiation) is 
  $4.5\times10^{14}$ D$_{900 pc}^2$ cm$^2$ in V2069 Cyg and $6.3\times 10^{13}$ D$_{1 kpc}^2$
  cm$^2$ in RX J0636. These values are still smaller than the size that
   we obtained for V2491 Cyg ($1.8\times10^{15}$ D$^2$$_{10.5 kpc}$ cm$^2$), but can be 
   comparable taking into account uncertainties in the distance determination for 
   both of them. 
   
  It should also be stressed that with the existing data, we cannot rule out that the supersoft
 flux has more than one origin, for instance that there are unresolved narrow emission
 lines merging with the soft continuum and making the source appear more luminous in
 the supersoft range. In archival data of V2487 Oph and V4743 Sgr \citep[][Zemko et al., in prep.]{her14V2487Oph} 
 we found that the post-outburst RGS spectra, albeit with low signal-to-noise ratio,
 still show emission lines in the softest range. At this stage, this possibility is only
 speculative, but it should be taken into account as it may imply a lower
 supersoft X-ray luminosity that estimated in our broad-band fit with {\sl Suzaku}.

 In calculating the luminosity we assumed the widely adopted distance 
 d=10.5 kpc \citep{hel08V2491cyg}. \citet{mun11v2491Cyg} independently found 14 kpc using 
 the interstellar \ion{Na}{1} line to evaluate the reddening. Although these 
 distances have been inferred with the 
 maximum-magnitude rate-of-decay (MMRD) relationship and these values may thus be highly 
 uncertain (the relationship does not hold for RNe and sometimes is not very precise for 
 CNe, moreover the nova had a rare secondary optical 
 maximum), the optical and especially the supersoft X-ray luminosity in the outburst indicate 
 that the distance cannot be much smaller. Especially the large supersoft X-ray flux 
 \citep{pag10V2491cyg, nes11V2491Cyg} is evidence of a distance of at least 10 kpc, otherwise 
 the WD would have been underluminous compared with the models of the same high effective 
 temperature \citep[from $9\times10^5$ to a $10^6$ K, see ][]{sta12, wol13Hburn} and 
 also with previous observations of novae in the SSS phase
 \citep[see][for a review]{ori12gratings}. 

\subsection{The 38 min period}
      
In our {\sl Suzaku} observations we detected the X-ray light curve modulation 
 with a period of 38.3 min (0.0266 days). \citet{shu10v2491cyg} detected a 0.02885 days (41.54 min)
  modulation in optical data in the $V$ band obtained between 2008 May 02 and 2009 November 23. 
We analysed the same dataset with the PDM method and found that the 0.02885 days period is most 
probably a daily alias of a 0.02655 days (38.23 min) period, which is very close to the one
we found in X-rays. Our analysis of the energy dependence of pulses indicates that absorption
 causes the observed 38.3 min modulation and the timescale implies that the absorber is 
 somewhere close to the WD. 
 The detected 38.3 min modulation may be the WD rotational period.
However, the peak that corresponds to this period in the LS periodogram is not as 
prominent as usually measured in IPs. There is a small, but puzzling difference in periods 
measured in the first and in the second halves of the {\sl Suzaku} exposure. However, given 
the data quality, we cannot rule out that the modulation is strictly periodic. 
 On the other hand, if the 38.3 min period is not related to the WD spin, it is not clear what 
 else may play a role in the time-dependent and compact absorber causing this modulation.   
 
\subsection{Magnetic driven accretion in V2491 Cyg}
Some indications of a magnetic nature of V2491 Cyg can be found in spectral characteristics, also 
discussed by \citet{tak11V2491Cyg}. These are our main findings:
\begin{itemize}
\item The hard X-ray band (2.0--10.0 keV) luminosity of V2491 Cyg
 is 1.82$\times10^{34}$ erg s$^{-1}$, which is higher than in most IPs (L$_{\rm{X}}$ 
 is usually $\lesssim 10^{33}$ erg s$^{-1}$  \citealt{war03CV, pre14IP}).  
In IPs the moderately hard X-ray emission mostly originates in high energy plasma produced by 
the accretion shock on the surface of the WD. 
\item The observed blackbody-like component may place V2491 Cyg in the group of ``soft IPs''.
The blackbody temperature and the emitting area are comparable with these systems.
\item If V2491 Cyg is a ``soft IP'', the soft X-ray flux emitting region underwent a seamless,
 early transition from nuclear burning to accretion spot emission, taking into account the 
 observations of \citet{pag10V2491cyg}.
\item There are two optically thin plasma components with characteristic 
temperatures $\sim$0.2 keV and $\sim$10--20 keV, which may be due to a temperature gradient in the 
post-shock region. 
\item We need a complex, partially covering absorber with N(H)$\sim10^{23}$ cm$^{-2}$ to
 fit the data, like in many IPs. This phenomenon is usually
  caused by the accretion curtains crossing the line of sight \citep{eva07softIP, anz08softIP}. 
\item The equivalent width of the Fe K$\alpha$ line in V2491 Cyg is very large, 246 eV, 
like usually in IPs \citep[see][for the Fe lines in IPs]{nor91FeKmCV, ezu99MCVASCA}. 
This is evidence of reflection of the X-ray emitting plasma, most probably from the surface
 of the WD \citep{geo91FeK}. It may also indicate 
copious X-ray emission above the {\sl Suzaku} XIS detectors' range. However, the observed 
Compton thin absorber can also contribute to the 6.4 keV line \citep[see][]{ezu99MCVASCA, ezeFe}.
\item The most widely accepted proof of an IP, in which the WD is not 
synchoronized with the orbital period, is the detection of the spin period in X-rays.
The X-ray flux is modulated because the accretion channeled to the pole is shocked, so it 
is self-occulted and partially absorbed as the WD rotates. 
We detected an X-ray flux modulation with a period of $\sim 38$ min that can be
attributed to the WD spin. Energy dependence of the amplitude of the pulses supports 
this assumption and is consistent with the accretion curtain scenario \citep{ros88exhyaEXOSAT, hel91aopsc}. 
\item However, this period is probably not entirely stable. On the other hand, detection of 
the spin period may not always be possible. \cite{ram08IP} suggested that if the magnetic and 
 rotational axes of a WD are closely aligned, the spin rotation may be undetectable.
\end{itemize}

\section{Conclusions}
We presented the {\sl Suzaku} XIS observation of the CN V2491 Cyg obtained in quiescence, 
more than two years after the outburst. The very high X-ray flux that we measured is close
 to the pre-outburst level \citep{iba09V2491cyg} and to the values that were obtained about
  250 days after the outburst \citep{pag10V2491cyg}. It indicates a high X-ray luminosity, 
  and hence high accretion rate. We found that the spectral characteristics of the
nova are very typical of ``soft IP'', such as the presence of a blackbody-like component with 
T$_{\rm{bb}}=77$ eV, optically-thin thermal plasma emission with two characteristic temperatures
and a partially-covering absorber. The hard band X-ray luminosity also indicates magnetic 
driven accretion. We detected the energy dependent X-ray flux modulation with 
period $\sim$ 38 min that may be attributed to the WD rotation. 
However, there are uncertainties in the stability of the 38 min period, and in the origin of the 
blackbody-like component which may be due to residual and localized nuclear burning.

 Magnetic WDs in novae have been observed in three systems: in two IPs (GK Per and DQ Her)
 and in one polar (V1500 Cyg). There are also several other IP candidates:
 V1425 Cyg, V533 Her and V842 Cen\footnote{
 The full catalog of IPs and IP candidates (Version 2014) by Koji Mukai, 
 http://asd.gsfc.nasa.gov/Koji.Mukai/iphome/iphome.html}, V4743 Sgr \citep{lei06V4743Sgr, dob10V4743Sgr}.
Little is known how the 
 magnetic field changes the nova evolution, it may really steer the path towards, or away 
 from, a type Ia SN explosion in RNe.

If V2491 Cyg is NOT an IP, more questions remain unresolved. The first concerns the possible
confinement of nuclear burning. A second question arises if accretion is channeled through
 the boundary layer of a disk; in this case the observed X-ray light curve 
modulation is most probably a quasi periodic oscillations, originating in the inner part of
 the accretion disk. However, this does not explain the energy dependence of the amplitude of the pulses.

An important measurement in further observations would be that of the orbital period. 
The {\sl Suzaku} observations, presented in this paper, lasted for slightly shorter than 
two days and we did not find the orbital modulation, usually observed in X-rays in IPs
\citep{par05IPorb}.

To conclude, V2491 Cyg seems to be a puzzling key object to understand the secular evolution
 of novae. We suggest that it should still be observed in different wavelengths. 
A hard X-ray observation with {\sl NuSTAR} would explain the origin of the strong 6.4 keV line. New soft 
X-ray observations would clarify the stability and nature of the $\sim$38.3 min modulation of the X-ray flux.
Finally, 
with optical observations we may measure the orbital period.

{\it Facilities:} \facility{Suzaku}.


\end{document}